# To Overhear or Not to Overhear: A Dilemma between Network Coding Gain and Energy Consumption in Multi-hop Wireless Networks

Nastooh Taheri Javan, Masoud Sabaei and Mehdi Dehghan
*{nastooh, sabaei and dehghan}@aut.ac.ir*
Computer and Information Technology Engineering Department
Amirkabir University of Technology (Tehran Polytechnic), Tehran, Iran

**Abstract**

Any properly designed network coding technique can result in increased throughput and reliability of multi-hop wireless networks by taking advantage of the broadcast nature of wireless medium. In many inter-flow network coding schemes nodes are encouraged to overhear neighbour's traffic in order to improve coding opportunities at the transmitter nodes. A study of these schemes reveal that some of the overheard packets are not useful for coding operation and thus this forced overhearing increases energy consumption dramatically. In this paper, we formulate network coding aware sleep/wakeup scheduling as a semi Markov decision process (SMDP) that leads to an optimal node operation. In the proposed solution for SMDP, the network nodes learn when to switch off their transceiver in order to conserve energy and when to stay awake to overhear some useful packets. One of the main challenges here is the delay in obtaining reward signals by nodes. We employ a modified Reinforcement Learning (RL) method based on continuous-time Q-learning to overcome this challenge in the learning process. Our simulation results confirm the optimality of the new methodology.

## 1. Introduction

The network coding fundamentals were proposed by Ahlswede et al. [1] in order to efficiently utilize resources of wired networks. With network coding, an intermediate node sends multiple packets within a single *coded* packet, so the most desirable feature of network coding is that it achieves significantly higher throughput via reducing the number of transmissions. Currently, opportunistic network coding is being applied to wireless networks, such as 802.11-based multi-hop wireless networks with substantial throughput improvements [2]. Network coding in the wireless networks can be classified into two types: when the coded packets are from different sessions, it is called *inter-flow* coding [3, 4], and when they are from the same session, it is called *intra-flow* network coding [5]. On the other hand, some network coding schemes, such as COPE [6], considered the *two-hop* coding frameworks (i.e. coded packets are decoded at the next hop of the coder node), whereas some others, such as DCAR [7] studied *multi-hop* coding frameworks (i.e. coded packets are decoded at two or more hops away from the coder node). In this paper, we explore two-hop inter-flow network coding in multi-hop wireless networks (similar to COPE).

Many network coding implementations in wireless networks encourage the nodes to overhear their neighbour traffic in order to provide increased coding opportunities and consequently reduce the number of transmissions. COPE [6] and DCAR [7] are two examples that are based on this idea. They demonstrate that overheard packets, in the context of broadcast wireless channel, can be efficiently used to assist network coding, resulting in transmission reductions and throughput improvements. For example, in COPE, the network nodes overhear the whole transmission traffic of their neighbors and keep these packets in their memory for a short period of time, this phase is known as *opportunistic listening* in the said scheme. In addition, in order to choose the best *coding pattern*, the coder node should be informed of the packets its neighboring nodes keep in their memory. Collecting this overhearing information at the coder node is possible through two approaches: one is by explicitly acknowledging, and the other is by statistical obtaining [6]. In the first method, a network node sends a list of the all the packets in its memory to its neighbors periodically as the *reception reports*, and in the second method, the overhearing information is guessed by the encoding node via using link quality advertisements through periodic probing [8]. Comparisons of different marking methods and their performance are beyond the scope of this paper.

In order to understand the importance of overhearing in network coding implementations, let us consider the simplified example in Fig. 1. In this scenario, there are two data flows, one from $n_1$ to $n_2$ (which includes packet $p_a$) and the other from $n_2$ to $n_3$ (which includes packet $p_b$). Suppose that in the first two transmissions, $n_1$ and $n_2$, respectively, send $p_a$ and $p_b$ packets to $R$ (as the intermediate node). Now $R$ shall send packet $p_a$ to $n_2$ and packet $p_b$ to $n_3$. Under this set up if $n_3$ does not try overhearing when $p_a$ is being sent by $n_1$, node $R$ has no coding opportunity for these packets and they shall be sent to their destinations through two separate transmissions. On the other hand, if $n_3$ does overhear $p_a$, $R$ can send the packets requested by $n_2$ and $n_3$ through coding $p_a$ and $p_b$ together, i.e. $p_a+p_b$ is sent in one transmission. This means that it requires one less transmission (3 rather than 4) to send $p_a$ and $p_b$ to their respective destinations. In general, in multi-hop wireless networks, if the network nodes overhear the neighbor's packets, they can improve *coding gain* (the ratio of the number of transmissions required by the non-coding approach to the number of transmissions used by a coding scheme).

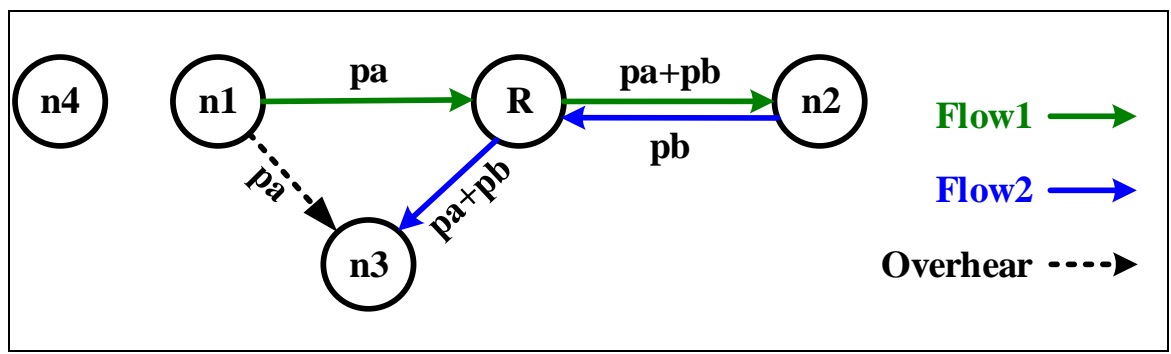

**Fig.1: The importance of overhearing in wireless network coding.**

It is important to note that the entire packets overheard by the nodes are not invariably useful for the coding operations. In other words, not all overheard packets boost coding opportunities. For example, overhearing packet $p_a$ in Fig. 1 by the other neighbors of $n_1$ (such as $n_4$) may not provide any coding opportunity at all, and therefore the worthiness of all overheard packets is not necessarily the same regarding coding gain.

On the other hand, wireless network nodes usually have finite power resource and how to improve network energy efficiency is still a main challenge. Wireless nodes have RF module that consume a considerable amount of energy for doing operations such as transmission, reception and idle listening. So, in practice, reducing energy consumption is a fundamental objective in wireless networks [9]. For example, in IEEE 802.11 [10], network nodes try to conserve energy through switching off the transceiver. In IEEE 802.11, in order to identify the receiver and also the transmission duration, the transmitter node broadcasts an RTS packet toward its neighbors, and then sends a data packet to the receiver node. The other nodes that have not been identified as the receiver turn off their transceiver for the transmission duration. We call these *idle periods* (i.e. overhearing opportunity). As mentioned before, unlike IEEE 802.11, in most network coding schemes network nodes are forced to stay awake over idle periods to overhear more packets.

Given this background and since all overheard packets are not always beneficial for coding operation, it is not desirable and cost effective for network nodes to remain awake all their idle periods in the hope of increased coding opportunities. Since nodes consume less energy when they switch to a sleep mode there is a trade-off between coding gain and energy consumption.

Some past works have tried to decrease overhearing overhead in wireless coded networks. For example, in [11], the study eliminated overhearing redundant packets in coded WSN through sending a new frame before each data packet to announce redundant packet at the receiver node. In [12] authors removed some of the overhearing slots by rearranging retransmission slots because they gave rise to some additional retransmission slots in their TDMA-based MAC protocol. To the best of our knowledge, most of the previous works have looked at the challenge and come up with restricted scenarios, whereas we aim to take this to a more general form in a COPE-like scenario in multi-hop wireless networks.

In this paper, we address the following question: Given the knowledge of received packets at the neighbors, how does one design an efficient sleep/wakeup scheduling for multi-hop wireless networks that can improve coding gain and energy consumption? The answer is to formulate the network coding aware sleep/wakeup scheduling as a semi Markov decision process (SMDP) [13], which is the approach adopted here which tries to predict the optimal wakeup and sleep patterns and determine optimal policy of the network nodes. We model the problem as SMDP rather than as Markov decision process (MDP) [14] because the time interval between two successive decision points is random. In discrete time MDP(s), the decision maker selects actions only at fixed epochs, however, in many practical problems, such as the one under study here, the times between the decision pints are not constant but random. In such scenarios an effective tool is SMDP which is consisted of states and actions, and is specified by the transitions probability and the sojourn time in each state when an action is executed.

In proposed scheme, at each idle period, the node decides to either: 1) "sleep," i.e. turn off it's transceiver to save energy; or 2) "stay awake" and thus overhear more packets. These idle periods can also be called *decision epochs*. Our approach enables the nodes in the multi-hop wireless networks to independently improve the overall energy consumption as well as coding gain. Using this SMDP formulation, we will investigate how nodes can learn the network dynamics, based on their local knowledge, and strive to find an optimal decision policy to maximize the cumulative rewards in long term.

We can solve SMDP problems using the classical method of dynamic programming (DP). However, DP needs all the transition probabilities, transition rewards and transition times to achieve results [14]. Instead, Reinforcement Learning (RL) [15] provides a convenient approach to solve decision-making problems for which optimal solutions theoretically are arduous and effortful to find, such as our case. What is important to note is that one of the main challenges with our SMDP is that there is stochastic time delay in the node's reward determination. In our problem, when a node decides to overhear a packet, usefulness or non-usefulness of the overheard packet is not determined immediately. Hence traditional reinforcement learning methods, such as Q-learning [16], are not applicable and a modified continuous-time Q-learning strategy for overcoming delayed rewards is called for. The results under simulation set up confirm that the proposed method provides a significant energy consumption improvement in multi-hop wireless networks.

Let us now briefly discuss what contributions this paper can make:

- We develop a foresighted decision making model to combine the sleep/wakeup scheduling and network coding frameworks. Because states of wireless networks change dynamically, foresighted nodes make optimal decisions on a long term basis. To enable the nodes to make suitable decisions to maximize long term outcome, we model the sequential decision-making as an SMDP at each independent node, so each node determines its best policy according to its local knowledge separately.
- We analyze trade-off between energy consumption and network coding gain. To do this, we explicitly explore the impact of sleep/wakeup scheduling on the network coding gain in wireless multi-hop networks.
- We propose a modified continuous-time Q-learning method for defeating the delayed reward problem in nodes' learning process.
- And finally, an optimal sleep/wakeup policy is derived by considering these two main aspects: the energy remaining in the node's battery and the network coding utility around the node.

The rest of the paper is organized as follows: Section 2 presents preliminary exploration of problem, Section 3 gives an overview of related works and system model is represented in Section 4. Then an analytical model is described in Section 5 before moving on to the next section to disclose our simulation results. The paper concludes in Section 7 recapping its main contributions.

## 2. Preliminary Exploration and Motivation

In order to understand the diversity of overheard packets in network coding, we implemented the COPE idea through NS2 simulator [17] employing one hundred nodes spread randomly over a 200×200 square meters. Each node had a 40 meters transmission range and each sent packet was 256 bytes in size and duration of each simulation was considered to be 300 seconds. During simulation, some CBR flows were randomly established between nodes with 12 seconds duration each, so that 14 flows on average were running in the network at each moment of simulation. DSR [18] algorithm and a modified slotted version of IEEE 802.11 were used for routing and MAC mechanism, respectively.

Under these conditions, nodes kept their transceiver switched on for overhearing during the entire period of simulation (as suggested in COPE). Results indicate that, on average, 38% of the packets overheard by the nodes provided no help to increase coding opportunities in the network. For example, Fig. 2 shows a total of 28 time slots of lifetime of a randomly selected node.

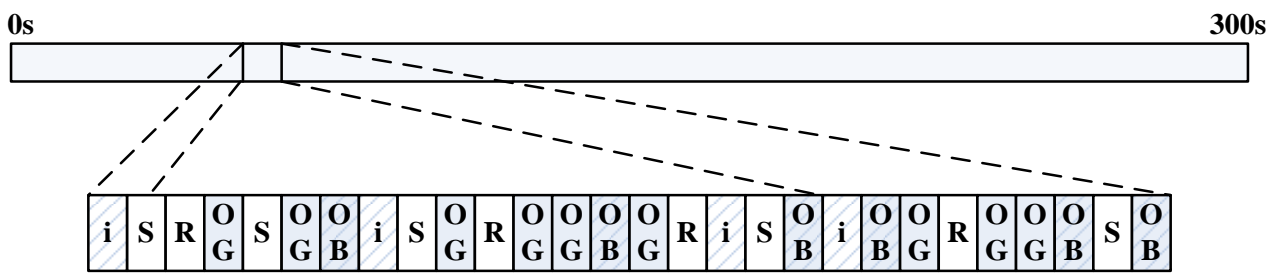

**Fig. 2. Narrative of a random node's life time in simulation.**

As can be seen from Fig. 2, each time slot may have five modes for each node:

1. Node is sending the packet (shown by white box denoted by letter S).
2. Node is receiving the packet (shown by white box denoted by letter R).
3. Node is overhearing a packet effective in coding (shown by gray box denoted by OG letters).
4. Node is overhearing a packet ineffective in coding (shown by hatched gray box denoted by OB letters).
5. Node is keeping the transceiver unit on, even though it does not receive any packet, i.e., idle listening. (Shown by hatched white box denoted by letter i).

Briefly, the results of our simulation demonstrate that during the entire time slots in which all the nodes kept their transceiver ON in the hope of overhearing useful packets, for 77% of the time the nodes overheard packets and for the remaining 23% they received no packet at all (*idle listening*). Furthermore, among the entire overheard packets, 38% did not help with the coding process. In other words, only during 48% of inactive periods the network nodes overheard helpful packets for network coding process and for the remaining 52%, during which the nodes deliberately kept their transceiver ON, awakening were not helpful (and this includes idle listening periods as well as receiving useless packets) and therefore it would be better for the nodes to turn OFF the transceiver over that periods.

To see what is involved in this process we ran our next experiments where parameters were set exactly as before and network topology and specifications of flows remained unchanged. In this new simulation scenario, the network nodes greedily tried to sleep, turning off their transceivers in all but the activity (send/receive) slots. In fact, in this case the nodes overheard no additional packet and tried to save their energy. Two clear conclusions can be drawn from these pair of investigations. Firstly, network energy consumption dramatically decreased in the second scenario resulting in 31% increase in its lifetime. (Here, network lifetime means the time interval between the beginning of simulation and the moment when the first node runs out of energy and dies). Secondly, coding gain in the first scenario is 22% higher (average coding gains in the first and second scenarios were measured as 1.38 and 1.13, respectively).

There is a trade-off between energy consumption and network coding gain in multi-hop wireless networks. Accordingly, we use an SMDP model to develop a strategy for the wireless node to decide its best course of action at every idle period. In practice, multi-hop wireless networks operate over stochastic environments with uncertainty and when faced with a sequential decision-making problem, a network node, as a decision maker, executes an action and transits from the current state to the next which affects the environment. Under these circumstances, some portion of the node's outcomes is random and some is under the control of the node. One of the most useful theoretical frameworks to configure these situations is SMDP which monitors their progress in a continuous-time manner. In Section 5-1 we will see how this problem can be formulated as a SMDP model.

We have already stated that the overheard packets by a node can help improve coding gain. After overhearing different packets a node declares it is storing packets in its memory to its neighbors and the neighbors may discover new coding opportunities upon this declaration. It is after these steps that usefulness or non-usefulness of an overheard packet for coding gain will be known at the moment of receiving a coded packet in several time slots later. Subsequently, whenever this node receives a coded packet containing the overheard packet, usefulness of the overheard packet will be revealed. In other words, when a node selects an action, the reward signals are not known immediately and may reach the node out of order. In order to overcome the delayed rewards challenge, we use a modified continuous-time Q-learning method that is described in Section 5-2.

## 3. Related Works

Reducing energy consumption in wireless communications has attracted increasing attention recently [9]. Many researchers study this from different perspective. For example, in [19] authors proposed energy efficient routing protocol for wireless environments and in [20], researchers strived to control energy usage through MAC techniques renovations. In [21, 22] some techniques came up with duty cycling approach and in [23], authors focused on topology control techniques to improve network life time. In alternative studies, some data driven techniques were reported to overcome energy dissipation. For example, authors in [24] argued that data aggregation could reduce energy expenditure, while in [25] the proposed solution demonstrated that nodes could improve network life time through data compression strategy and in [26] authors achieved this goal via a network coding scheme. Among these techniques, network coding improves energy consumption through decreasing the number of transmissions in wireless networks. For example, Cui et al. proposed CORP by using a suboptimal scheduling algorithm that exploited network coding opportunities [27], and thus achieved a significant power saving over pure routing for multiple unicast sessions. Also, a simple and low complexity subgraph selection method is revealed in [28] that reduces energy consumption in coded wireless networks.

Efforts have also been made lately to reduce overhearing in restricted scenarios (e.g., [11, 12]) and to exploit incomplete overhearing (e.g., [30, 31 and 32]) in wireless network coding. In [11], authors eliminated overhearing redundant broadcast packets in coded WSN by using a new small frame, which they called *digest info*. They proposed that digest info packet should be sent right before each broadcast data packet. The digest info packet has a field that contains either a unique identifier, or a hash of the data contained in the subsequent data packet. The node uses the information in the digest info to identify the subsequent data packet. If a node finds that the following data packet has already been received, it can put its radio to sleep mode. In [12] authors applied network coding in GinMAC (a TDMA-based MAC protocol [29]). They managed to eliminate some overhearing slots by rearranging retransmission slots (in practice by adding more retransmission slots!).

In [30] authors disclosed a network coding scheme adapted to a two-way relay (*reverse carpooling*) scenario with fading and noise corruption in which overhearing was often imperfect. They believed it was clearly wasteful to ignore the whole overheard information, given that only a few symbols in it were incorrect. So they divided the overheard packet at the destinations into two parts: one part with high probability to be correct (clean overheard symbols); the other with high probability to be incorrect (faulty overheard symbols). Eventually, at the destination, the first attempt tries to recognize the interference via next packet and the second attempt tries to reconstruct faulty signals by removing known interference. In [31] a NC scheme was introduced which actively exploited the overhearing channels by adapting the transmission rate to both the direct-channel and overhearing-channel conditions. In this scheme the opportunistic listeners receives data at a lower rate. In addition, the receiver can only overhear a part of the data-flow by rate splitting technique at the sender node. So, in poor overhearing-channel conditions the throughput can improve. In [32] authors investigated how the coding node can be efficiently informed of the content of the decoding buffers of the receivers. They studied a deterministic system, where the overhearing packets were announced to the coding node via reports, or via stochastic system (where the coding node makes decisions based on statistics without explicit reports). Their work demonstrated that maximum throughput can be achieved without explicit ACKs, i.e. by utilizing scheduling algorithms that guessed, encoded and then corrected the transmissions through feedback. In particular, they concluded that when the overhearing probability at the receiver was 1 (perfect overhearing) their scheme had no loss of optimality and that for probability values above 0.6 the loss of throughput in comparison to the explicit ACKs was very small (i.e. less than 5%).

It is worth noting that a recent research has been reported that improves the multi-hop network coding performance through virtual overhearing [33]. This method allows a node to obtain the packet sent by another node that is multiple hops away for free. An important point to bear in mind is that three conditions (regarding nodes and flows [34]) must be satisfied in order to establish virtual overhearing. Once, virtual overhearing has been established it enables a node to obtain the packets from another node and thus they can be used to increase coding opportunities.

From solution perspective, the studies on incorporating network coding and duty cycling also touch on our work at the first glance, but they were proposed to achieve different goals. For example, an efficient communication paradigm was proposed in [35] that applied network coding to bottleneck zone around the sink node in duty cycled WSN. Authors believed the nodes in bottleneck zone consumed more energy compared to other nodes due to heavy traffic. They divided the sensor nodes in the bottleneck zone into two groups: simple nodes and coder nodes and discovered that energy efficiency of the bottleneck zone increased due to coder nodes operation resulting in higher amount of data (load) to be transmitted to the sink for equal number of transmissions. Similarly, in [36] authors unveiled a communication scheme which employed network coding and duty cycling for bottleneck zone in multimedia WSN. In [37 and 38] DutyCode combines the idea of network coding and duty cycle in the MAC layer only in flood-based WSN. The main idea in this study is to exploit the redundancy in flooding applications (which uses network coding), and to put the node to sleep mode when a redundant transmission takes place. In particular, DutyCode tries to put a node in sleep mode when this node has already received and successfully decoded a sequence of coded packets. In [39] GreenCode was put forward as an energy efficient network coding aware MAC protocol which employed bi-directional transmission (i.e. reverse carpooling scenario).

## 4. System Model

In this section system model is presented briefly. We consider a stationary multi-hop wireless network [40], supporting multiple unicast sessions. The wireless network topology, given by the nodes and the links corresponding to pairs of nodes within direct communication range, is as a graph $G = (N, H)$ with node set $N$ and directed edge set $H$. Each node in the network can be a source or destination of traffic. Network nodes are uniformly and independently distributed over a two-dimensional region and communicate only with other nodes within a neighbourhood of radius $\rho$. A homogenous setup is considered for network nodes, i.e., we assume all nodes support the same maximum transmission rate, say 11 Mbit/s, and that all have omni-directional antennas. These wireless nodes operate in half-duplex mode and communicate with each other and share the common channel bandwidth. Each node can overhear all the transmissions from its one-hop neighbors. Additionally, our analysis assumes that all packets have the same size, and that the required overhead of transmitting coding coefficients is negligible.

Nodes have a transmission queue for pure packets to be sent to the neighbors. The number of existing packets in this queue is denoted as $p$. Nodes put the pure packets which they have overheard opportunistically in a separate queue called overhearing queue. The number of packets existing in this queue is denoted as $q$. Overheard packets all have the same finite life time that once it is over they will be omitted through *aging* algorithm. In the analytical model, we assume that there is no repeated packet in the network (since repeated packets would be detected and omitted by other layers), that each node is energised by a battery and a node dies once its battery is flat. Since battery replacement is either impossible or expensive in various applications for wireless nodes [41], we assume that the network nodes can use one of the energy harvesting methods [42]. Let's also suppose that a node consumes $E_T$ units of energy during each packet transmission and $E_R$ units of energy when receives or overhears a packet. In each time slot when a node does not receive any packet (idle listening) it loses $E_I$ units of energy, and finally that the amount of energy consumed in sleep mode can be ignored. We now focus on the inter-flow coding fashion similar to the ones used in COPE [6]. In the XOR coding, a coding node is one which encodes packets for several flows and that coding flows are the flows that are transmitted via a coding node and their packets have the opportunity to be encoded. Any node that receives an encoded packet is able to decode it using the unencoded, or native packets captured from the wireless channel. The summary of notations is presented in Table 1.

## 5. Analytical Models

In this section, we describe the analytical models in details. Section 5-1 covers the SMDP formulation and learning process is investigated in section 5-2.

Published in *Wireless Networks*, Springer, 2018. DOI: 10.1007/s11276-018-1733-0

**Table 1: Summary of notations.**

| | | | |
|---|---|---|---|
| $A(s)$ | set of all possible actions in state $s$ | $Q$ | Q value entity |
| $a^*$ | most valuable action | $q$ | transmission queue length |
| $a_t$ | taken action in time $t$ | $P$ | transmission radius of each node |
| $a_{t+1}$ | next action | $s_t$ | state in time $t$ |
| $\alpha$ | usefulness deadline of an overheard packet | $s'$, $s_{t+1}$ | next state |
| $d_t$ | $t^{th}$ decision epoch | $T$ | time index |
| $E_I$ | idle listening energy consumption | $T_{slot}$ | duration of each packet transmission |
| $E_R$ | receiving energy consumption | $W_{rc}$ | transmission power |
| $E_T$ | transmission energy consumption | $W_{tr}$ | receiving power |
| $e_t$ | residual energy in time $t$ | $Y_{dd'}$ | expected time between decision epochs $d$ and $d'$ |
| $F_s(a)$ | sojourn time in state $s$ by action $a$ | $B$ | learning rate |
| $p_t^c$ | received coded packet in time $t$ | $\Gamma$ | discount factor in learning process |
| $p_t^o$ | overheard packet in time $t$ | $\theta_{max}$ | maximum feasible value of $\theta$ |
| $g_t$ | average coding degree of packets in time $t$ | $\hat{\theta}$ | delay estimate |
| $K_n$ | set of neighbors of node $n$ | $\hat{\theta}^*$ | most valuable delay estimate |
| $p$ | overhearing queue length | $\lambda_{ij}$ | intensity of transmission packet from $i$ to $j$ |
| $P_{xy}$ | transition probability from state $x$ to $y$ | $\psi$ | usefulness factor of overheard packet |

**5-1- Semi Markov Decision Process Model**

In general, MDP provides an analytical structure to model sequential decision-making problems in circumstances where some portion of results is random and some is under the control of a decision maker. A decision maker at every iteration in discrete-time MDP selects an action which can affect the stochastic system, i.e. both the next system state and the reward are dependent on the executed action. In these systems, the final goal of the decision maker is to execute a chain of actions that maximizes cumulative revenue at the end of a finite or an infinite horizon. Finding an optimal sequence of actions is the solution of the MDP problem [14]. To do this, the decision maker considers the future states and correlated rewards in action selection process at the current state.

Instead of discrete-time MDP, time is considered as a continuous parameter in SMDP. In SMDP, the decisions can be selected at the instant when any random event happens, so SMDPs allow time spent in a particular state to follow an arbitrary probability distribution [13]. In this model, the system state may change several times between decision epochs, unlike MDPs where state changes are only due to actions. An SMDP is defined as a five tuple ($S$, $A$, $P$, $R$, $F$), where $S$ is a finite set of states, $A$ is a set of actions, $P$ is a set of transition probabilities, $R$ is the reward function, and $F$ is a function giving expected sojourn times for each state-action pair [13].

We formulate our problem as SMDP in which the time interval between two successive decision epochs is continuous random variable. The goal is to develop a strategy for the wireless node to select its best action at every idle period and optimize the total discounted reward. In the proposed model, the network nodes (as decision makers) make an effort to learn when to stay awake and when to sleep in their idle periods independently. Detailed formulation of the system will be described in the following sections.

**Decision Epochs**: Decision epochs are a sequence of time points $\{d_0, d_1, d_2, \ldots, d_t, \ldots\}$ at which an action is selected and state transition may happen. In our problem the decision epochs are the arrival instances of the idle periods (i.e. opportunity of overhearing). During the node life cycle, the arrival instances of the idle periods are random and the arrival instance can be viewed as a random sequence of epochs along time. Thus, at each available overhearing opportunity, the node decides to either overhear or sleep. These available overhearing opportunities can also be called decision epochs. The distribution of the inter-arrival time of the decision epochs could be arbitrary and depending on the neighbors' topology and local traffic.

**State space:** We denote the set of possible node states by $S$ in which $S$ is a finite set. Let $s_t \in S$ denote the state of the node at time $t$. The system state $s_t$ at time $t$ is given by:

$$s_t = (e_t, g_t) \qquad (1)$$

where $e_t$ is residual node's energy and $g_t$ indicates the estimated value of the network coding utility in the node's neighbourhood in time $t$, both $e_t$ and $g_t$ should be quantized into discrete intervals. Each node initially starts with a full battery capacity of $E$, we denote $e \in E = \{1, 2, \ldots, E\}$ as the residual energy state space. Also, the node estimates $g_t$ by a simple algorithm that calculates the average *coding degree* (the number of packets combined in one encoded packet) of the last $c$ received packets. The higher values of $g_t$ means that much coding operations has taken place in recent past around the node. The $c$ factor can be determined during the implementation phase.

**Actions:** Depending on system state $s$, and in each decision epoch, the network node determines to overhear or to sleep. Let $A(s_t)$ denote the set of all possible actions in state $s_t$ and $a_t$ be the control action executed at time $t$. Each action in $A$ corresponds to the following values:

$$a_t = \begin{cases} 0, & overhear \\ 1, & sleep \end{cases} \qquad (2)$$

**State dynamics:** The state dynamics of the nodes can be determined by the state transition probabilities, $P_{ss'}(a)$, and the expected sojourn time, $F_s(a)$, for each state-action pair. $P_{ss'}(a)$ can be defined as the probability that at the next decision epoch the system will be in state $s'$ if action $a$ is selected at the present state $s$ while $F_s(a)$ is the expected time until the next decision epoch after action $a$ is chosen at the current state $s$. Also, We defined $Y_{dd'}$ as expected time between two successive decision epochs $d$ and $d'$. $Y_{dd'}$ depends on different stochastic parameters, such as traffic pattern, local topology and so on. Stated simply, $Y_{dd'}$ is the time duration until the next overhearing opportunity. On the other hand, overhearing opportunities occurs for node $n$ when its neighbors transmit a packet to their neighbors except the node $n$. Let us assume each node transmits packets to each neighbor according to mutually independent Poisson processes and let $\lambda_{ij}$ denote the intensity of transmission packets from node $i$ to node $j$. The cumulative overhearing opportunity rate is the sum of the rates of all constituent processes, and thus the expected $Y_{dd'}$ for the node $n$ is the inverse of the event rate:

$$Y_{dd'} = \left[\sum_{i \in K_n} \sum_{j \in (K_i - n)} \lambda_{ij}\right]^{-1} \quad (3)$$

where $K_n$ is set of neighbors of node $n$. In equation (3), the cumulative overhearing opportunity for the node $n$ is calculated by summation of the intensity of transmission packets from all the $n$'s neighbors to their respective neighbors except $n$, where $\lambda_{ij}$ indicates the intensity of transmission packets from node $i$ to node $j$.

**Optimal policy:** A policy $\pi = \{(s,a) \mid a \in A, s \in S\}$ is a set of state-action pairs for all states of an SMDP. An optimal policy is the one maximizing the expected cumulative reward.

**Reward function:** The main goal of network coding in wireless networks is to reduce the number of transmissions and consequently to save energy. Without loss of generality, we define a node's revenue function based on the amount of energy conservation on the whole of the network due to its action. In particular, our aim is to design a sleep/wakeup scheduling algorithm for a node's idle periods that minimizes the total network power expenditure for overhearing while maintaining network coding gain. It is worth noting that the main source of energy consumption in wireless network is the transceiver unit due to activities such as data transmission, data reception, overhearing and so on.

The energy consumed at the node $n$ by the data transmission operation for the duration of one time slot, $T_{slot}$ is:

$$E_T = W_{tr} \times T_{slot} \quad (4)$$

where $W_{tr}$ is a constant amount of power when the node $n$ is in transmission mode. We assume that all $T_{slot}$s have the same duration which is long enough for transmitting one packet, so each send or receive operation occurs in one $T_{slot}$. Also, each network node consumes energy while receiving the data packet transmitted by one of its neighbors. We assume the transceiver at each node consumes a constant amount of power $W_{rc}$ when in receive mode. Hence, the energy consumed for packet reception during time slot $T_{slot}$ is:

$$E_R = W_{rc} \times T_{slot} \quad (5)$$

We assume also that the transceiver does not consume any energy when the node is in sleep mode. As a matter of fact and based on the above assumption, if the node sleeps over one idle period it can save energy equal to $E_R$.

The real-valued function $r$ denotes the value of the reward received at time $t$. The reward is referred to as income or cost depending on whether or not $r$ is positive or negative, respectively. The node's reward function can be defined as:

$$r(s_t, a_t) = Revenue(s_t, a_t) - Cost(s_t, a_t) \quad (6)$$

where:

$$Revenue(s_t, a_t) = \begin{cases} 0, & a_t = 1 \\ \psi E_{tr}, & a_t = 0 \text{ and } p_t^o = useful \\ 0, & a_t = 0 \text{ and } p_t^o = useless \end{cases} \quad (7)$$

and

$$Cost(s_t, a_t) = \begin{cases} 0, & a_t = 1 \\ E_{rc}, & a_t = 0 \end{cases} \quad (8)$$

where $\psi (0 \leq \psi)$ is usefulness factor of overheard packet (i.e. frequency of participation of overheard packet in neighbors' coding operation) and $p_t^o$ is overheard packet in time $t$, so we have:

$$p_t^o = \begin{cases} useful, & if\ p_t^o\ is\ in\ p_k^c \\ useless, & else \end{cases} \quad \exists\ p_k^c,\ t \leq k \leq t + \alpha \quad (9)$$

where $p_k^c$ is the received coded packet in time $k$, $t$ is the time that packet $p_t^o$ was overheard and $\alpha$ is usefulness deadline of $p_t^o$ (in other word, $\alpha$ is the duration that the node keeps the overheard packet in its buffer). Each useful overheard packet participating in one coded packet reduces exactly one transmission in the network. Of course, an overhead packet may be used many times over in neighbors' coding operation, as result more than one transmission would be saved. When an overheard packet participates in $\psi$ coded packets, it reduces $\psi$ transmissions in the network. This means if an overheard packet is useless, the $\psi$ factor for this packet is equal to zero. Intuitively, the *Revenue* function's value indicates the amount of saved energy in the node's neighbourhood.

**Remark1:** We cannot assume that the agent will receive rewards immediately after performing $a_t=0$. In particular, when a node chooses $a_t=0$ action (i.e. overhears a packet) in decision epoch $d_t$, usefulness or non-usefulness of the received packet is not known immediately because the node has to send the reception report to all of its neighbors and they may not use the overheard packets in their coding operation. However, the overheard packet becomes useful when neighbors use it in their coding operation and the node will know of this when it receives a coded packet that includes the overheard packet. Therefore, pure Q-learning [16] is not applicable to our problem and what is needed is a modified continuous-time Q-learning method to dominate delayed rewards. We will discuss this in greater details in the Section 5-2.

Fig. 3 illustrates a schematic view of events in our problem. As can be seen there are 5 decision epochs ($d_{t-2}$ to $d_{t+2}$) that runs for a given node. The node has chosen action $a=0$ in decision epochs $d_{t-2}$, $d_{t-1}$ and $d_{t+1}$ and received overheard packets $p_{t-2}^o$, $p_{t-1}^o$ and $p_{t+1}^o$, respectively. In fact, when the node received these overheard packets, it didn't know which packet was useful for neighbors' coding operation and which was useless. In addition, the node received some coded or native packets from its neighbors (only coded packets are seen in Fig. 3 as $p^c$s). Suppose that after some time the node received $p_3^c$ from one of its neighbors which included $p_{t-2}^o$ and found the packet $p_{t-2}^o$ was useful in coding operation for its neighbor. Hence action $a=0$ in decision epoch $d_{t-2}$ was the right choice. Now let us look at the destiny of $p_{t-1}^o$ that was overheard in $d_{t-1}$. Unfortunately, after $\alpha$ times of $d_{t-1}$, the node received no coded packet that included $p_{t-1}^o$, so the node concluded that action $a=0$ in time $d_{t-1}$ was unhelpful.

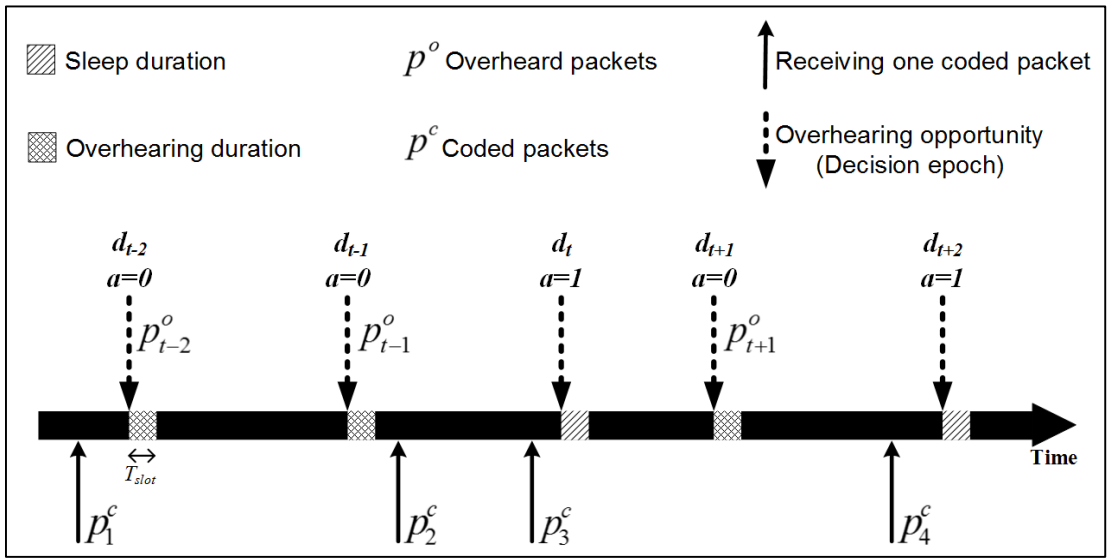

**Fig. 3.** Schematic depiction of the decision model for overhearing at a node.

**5-2-Learning Model**

There are several algorithms that offer optimal solutions for an SMDP, such as policy iteration, value iteration and linear programming [13], however, these model-based solutions need past information of system, such as previous state transitions and transition rewards. In many practical scenarios, as in our case, the transition probability $P_{ss'}(a)$ and sojourn time $F_s(a)$ are either unknown or difficult to express in close form which makes it hard to obtain an optimal policy. Instead of these approaches, Reinforcement Learning (RL) presents an appropriate algorithm for decision-making problems for which optimal solutions are onerous to define theoretically. Q-learning [16] is a favorite and well-known approach to learn from reinforcements. In Q-learning, a decision maker tries to obtain an optimal policy without any prior information about the transition times and probabilities.

In discrete-time version of Q-learning, value function is demonstrated by a two-dimensional table indexed by state-action pairs. In particular, for each state *s* and action *a*, Q value is defined as:

$$Q(s_t, a_t) = R(s_t, a_t) + \gamma \sum_{s_{t+1} \in S} P_{s_t s_{t+1}}(a) Q(s_{t+1}, a_{t+1}) \tag{10}$$

where *γ* is discount factor (0≤*γ*<1). Intuitively, *Q(s,a)* is the maximum discounted cumulative reward that can be gained starting from state *s* and executing action *a* at the first step. From above definition, $Q^*$ satisfies Bellman-style optimality equation:

$$Q^*(s_t, a_t) = R(s_t, a_t) + \gamma \sum_{s_{t+1} \in S} P_{s_t s_{t+1}}(a) \max_{a_{t+1} \in A(s_{t+1})} Q^*(s_{t+1}, a_{t+1}) \tag{11}$$

A decision maker updates Q-values during learning process to estimate $Q^*$ by value iteration method based on following rule:

$$Q(s_t, a_t) \xleftarrow{update} Q(s_t, a_t) + \beta \times \left( R(s_{t+1}, a_{t+1}) + \gamma \max_{a_{t+1} \in A(s_{t+1})} \{Q(s_{t+1}, a_{t+1})\} - Q(s_t, a_t) \right) \quad (12)$$

where constant $\beta$ controls the learning rate.

As mentioned before, SMDP is continuous-time generalization of discrete-time MDP, and system progress is modeled in continuous time fashion in SMDP, therefore a continuous-time version of Q-learning [43, 44] should be used for solving SMDP. In this case, we update Q-value at decision epoch $d_{t+1}$ according to the following equation:

$$Q(s_t, a_t) = Q(s_t, a_t) + \beta \times \left( R(s_{t+1}, a_{t+1}) + e^{-\gamma(d_{t+1} - d_t)} \max_{a_{t+1} \in A(s_{t+1})} \{Q(s_{t+1}, a_{t+1})\} - Q(s_t, a_t) \right) \quad (13)$$

In general applications, researchers assume that when a decision maker selects an action, it receives reward (or cost) immediately. However, in some applications, such as ours, it is possible that reception of rewards is delayed in time. This kind of delay is due to many possible reasons, e.g. it may be the result of unpredictable natural latency in these applications.

In our case, when action $a_t=0$ is selected and a packet is overheard, reward is not known immediately and the node has to wait until the effect of the overheard packet is disclosed. The reason for this is that the node has to propagate the reception report, including the list of overheard packets, to all its neighbors and then receive coded packet as to whether or not the overheard packet was useful in their coding operation. Only then does the node know if choosing action $a_t=0$ was beneficial and qualifies for reward. This reward is proportionate to coding degree of received coded packet. Based on these assumptions rewards may reach the node out of order. (see Fig. 3 and related description in Section 5-1).

Let us now examine the literature which has extensively addressed the problems of constant and stochastic time delays in rewards. For example, authors in [45] investigated constant delays and reduced their problem to a MDP through expanding state space and redefining cost function. In [46] the study assumed that all reward signals arrived in the correct order and they just moved in time unequal. In fact, the authors assumed that the number of time steps between rewards is a non-negative random variable. These two studies demonstrated that if decision maker has accurate knowledge about the delay, it can assign reward (and cost) to correct state-action, and so its problem is reduced to a normal MDP.

In [47] Campbell et al. made the assumption that each reward is independently delayed by a non-negative random variable in discrete-time MDPs. This meant that rewards might reach the decision maker out of order. They further assumed that the amount of delay was equal to a random Poissonian variable with mean and variance $\theta$. They extended discrete-time Q-learning to comprise a delay dimension so that the algorithm could measure the value of state-action-delay.

We were inspired by Campbell's method [47] and on the basis of their work we propose a continuous-time Q-learning method. Let us assume that all reward values will be delayed in time by a Poissonian random variable in terms of decision epochs. In order to resolve the problem arising from delayed rewards, we decided to introduce delay into the learning process. This was done by adding a new field to the Q-values which accounted for the delay value elements. Each entry in the Q-table can be represented by Q($s, a, \hat{\theta}$), where $\hat{\theta}$ is a discreet variable which indicates delay estimate according to decision epochs and $0 \leq \hat{\theta} \leq \theta_{max}$, where $\theta_{max}$ is the maximum possible value of $\theta$ and $\hat{\theta}$ enumerates the decision epochs for delay estimate. In this case, the Q-table space gets bigger because of the acquired additional dimension $\theta$, but still due to small space of both actions and states the extended version is acceptable in practice. It is worth noting that the Poisson distribution has no maximum limit theoretically but because of implementation restrictions we have to limit it to $\theta_{max}$ in order to obtain an acceptable space for Q-table. The delay is not expected to be greater than $\theta_{max}$ in practice, however, if it exceeded $\theta_{max}$, the effect on performance would still be unimportant.

Intuitively, in our learning method, the node should learn about the delay. To do this, the node must store sufficient number of its previous operations in its memory in order to consider delayed rewards in learning process. In fact, the node maintains multiple Q-values per each possible value of time delay in terms of decision epoch instances. What is important to note is that in each iteration all Q-values related to current state-action pair should be updated concurrently (for all values of $\hat{\theta}$). The most valuable $\hat{\theta}$ for each state-action pair will tend to the true value of $\theta$ in the long term, i.e., during learning process, the node tries to find the true value of delay estimate. In other words, the node maintains all possible values of delay and tries to use the most useful delay estimate in every decision epoch. It is predicted that some estimates will get more credit than others.

In our learning framework, the new action space size is equal to $|A| \times (\theta_{max}+1)$ for each state and the node should select not only an action but also a delay estimate. At each decision epoch, the node tries to operate according to the best delay estimate, and so at first it select the most valuable delay estimate $\hat{\theta}^*$ where:

$$\hat{\theta}^* = \arg\max_{\hat{\theta}} Q(s, a, \hat{\theta}) \tag{14}$$

Then, on the assumption that this is an accurate estimate the node selects the entire Q-values, with $\hat{\theta}$ being equal to $\hat{\theta}^*$, and picks the most valuable action among these selected Q-values according to:

$$a^* = \arg\max_{a} Q(s, a, \hat{\theta}^*) \tag{15}$$

Next, the node executes this action and then moves on to state $s'$. At this point, the node selects the next estimate $\hat{\theta}^*$ and the operation repeats itself until exploration of the environment is completely exhausted. Fig 4 describes this process schematically. In Fig 4, the most valuable delay estimate (θ* in the figure) was selected and the node should select the best action from the selected two-dimensional Q-table (the gray one).

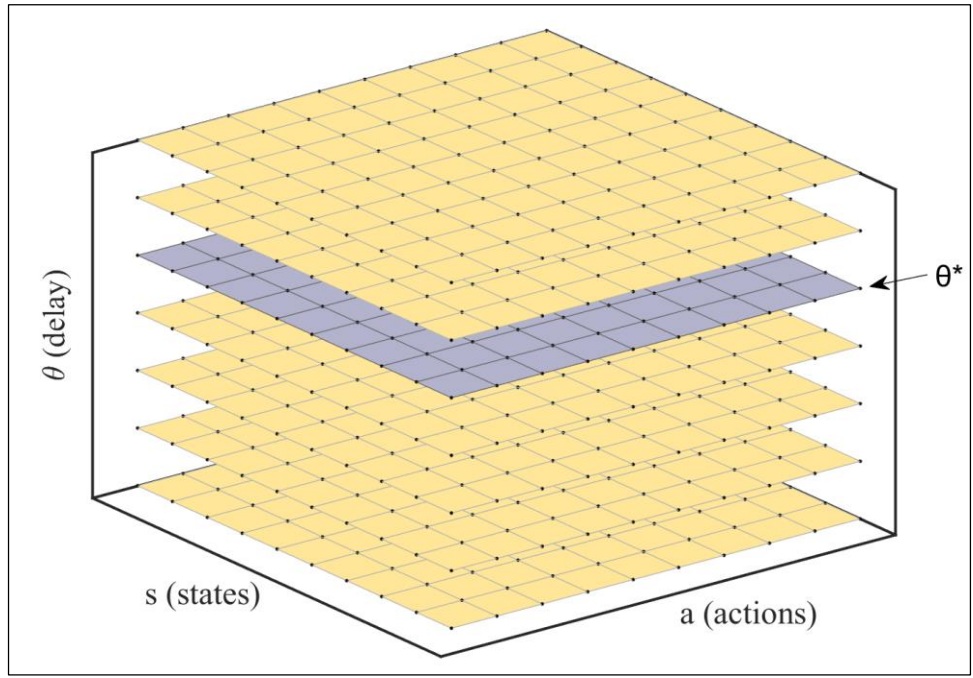

**Fig. 4. the valuable delay selection.**

Based on our assumptions, this node will receive the reward belonging to this action after the random delay, but instead, it may receive a reward related to past actions during this time step. It is possible that the node earns two (or more) rewards simultaneously which occurs when it receives a coded packet that includes two (or more) overheard packets. Also, it may gain a reward according to a specific action in two (or more) parts, when the overheard packet participates in two (or more) codec packets (i.e. $\psi \geq 2$ for this overheard packet). Actually, during learning process, the node does not know the true value of the mean delay, so it allocates credit to all the Q-values concurrently. The node updates all the *Q*-value estimates in each round and assigns the reward to each of them at all estimated value of delay. To do this, it updates the Q-values for all $\hat{\theta} \in |\Theta|$ based on the following expression:

$$Q(s_{t-\hat{\theta}}, a_{t-\hat{\theta}}, \hat{\theta}) \xleftarrow{update} Q(s_{t-\hat{\theta}}, a_{t-\hat{\theta}}, \hat{\theta})$$
$$+ \beta \times \left( R(s_t, a_t) + e^{-\gamma(d_{(t+1)-\hat{\theta}} - d_{t-\hat{\theta}})} \max_{a_{t+1} \in A(s_{t+1}), \hat{\theta}} \{ Q(s_{(t+1)-\hat{\theta}}, a_{(t+1)-\hat{\theta}}, \hat{\theta}) \} \right.$$
$$\left. - Q(s_{t-\hat{\theta}}, a_{t-\hat{\theta}}) \right)$$
(16)

where $\hat{\theta}$ is the $\hat{\theta}^{th}$ estimate of $|\Theta|$ total estimates.

In order to explain the updating method let us take an example where a node tries to update Q(*s*, *a*, 3), it supposes that the delay is equal to three decision epochs, which means that all credit should be assigned to the state-action pair three decision epochs earlier, because if the delay is actually three decision epochs long, then this state-action pair must have been the cause. In the same way, the node updates the other Q-values and over time, some estimates will get much more value than others. The details of the learning algorithm are given in Algorithm 1.

| | |
|---|---|
| **Algorithm 1: Q-learning based learning algorithm** | |
| 1: | **Initialize** the following parameters: |
| | • maximum feasible delay: $\theta_{max}$ |
| | • learning rate: $\beta$ |
| | • discount factor: $\gamma$ |
| | • Q values: $Q(s,a,\theta)$ |
| | • initiate state: $s_0$ |
| 2: | **Repeat** |
| 3: | select $\hat{\theta}^*$ according to $\hat{\theta}^* = \arg\max_{\hat{\theta}} Q(s,a,\hat{\theta})$ |
| 4: | suppose $\hat{\theta}^*$ is correct and then select all Q-values with $\hat{\theta}$ being equal to $\hat{\theta}^*$ |
| 5: | select $a^*$ among selected Q-values in step 4 |
| 6: | apply action $a^*$ |
| 7: | observe $R$(rewards) and $s_{t+1}$(next state) |
| 8: | **for** $\hat{\theta} := 0$ to $\theta_{max}$ do |
| 9: | $$Q(s_{t-\hat{\theta}}, a_{t-\hat{\theta}}, \hat{\theta}) \xleftarrow{update} Q(s_{t-\hat{\theta}}, a_{t-\hat{\theta}}, \hat{\theta}) + \beta \times \left( R(s_t, a_t) + e^{-\gamma(d_{(t+1)-\hat{\theta}} - d_{t-\hat{\theta}})} \max_{a_{t+1} \in A(s_{t+1}), \hat{\theta}} \{Q(s_{(t+1)-\hat{\theta}}, a_{(t+1)-\hat{\theta}}, \hat{\theta})\} - Q(s_{t-\hat{\theta}}, a_{t-\hat{\theta}}) \right)$$ |
| 10: | **end** for |
| 11: | $s=s_{t+1}$ |
| 12: | **until** end of learning process |

This method can be viewed as a special case of Q-learning and the convergence proofs for Q-learning [48] can be extended to the case of this method [49].

## 6. Numerical Results and Discussion

In this section, our focus will be on simulation operation that aims to evaluate the proposed scheme. In Section 6.1, we explain the simulation setup and the various simulation parameters and in Section 6.2, we present simulation results.

**6.1. Simulation Environment**

The experiments were performed using the NS2 [17] based simulation framework which is common in wireless network research. We implemented both COPE [6] and our scheme in our evaluation process. We considered 200 static nodes that were deployed in an 1100×1100m$^2$ square field and signal attenuation was modeled by the two-ray ground propagation model [50]. We implemented IEEE 802.11 as MAC protocol and used a very simple geographic routing, so the routing protocol picked the neighbor closest to the destination as the next hop.

Next, we employed UDP traffic, where the packet size was set to 500B. UDP traffic enabled us to clearly demonstrate the superiority of our network coding solution at the MAC layer and so it was unnecessary to consider TCP at the transport layer. We changed the network offered load by manipulating the number of flows in the network. Flows were established randomly between two nodes as source and destination. Traffic load was generated on all sources with the same intensity using exponential distribution of inter-arrival times.

Also in this set up, we applied a power consumption model similar to the model in [51], as shown in Table 4. Nodes had a nominal transmission range of $\rho$=200m, hence all the nodes within this range could receive the transmission. We set learning rate $\beta=0.5$, discount factor $\gamma=0.9$ and $\theta_{max}=8$ and results were obtained by taking average over 250 runs, each of which ended after 500 seconds.

Moreover, in order to reduce the state space we quantized $e_t$ and $g_t$ into discrete intervals. In this way, we considered $e_t$ (i.e. node's level of residual power at time $t$) as a variable that could have discrete values from one to eight and it was assumed the energy storage capacity of a node to be 8 units. Similarly, the variable $g_t$ (i.e. average coding degree of received packets) taking discrete values ranging from one to ten. If the average coding degree of the last 15 received packets (i.e. $c$ was set to 15) was between 1 to 1.1, the value of $g_t$ was 1, if that was between 1.1 to 1.2, $g_t$ moved up to 2, and so on. Finally when coding degree was between 1.9 to 2, $g_t$ stood at 10.

Three important performance metrics that were evaluated included: (i) Average coding gain, (ii) Average end-to-end delay (iii) Average energy consumption per bit, and the parameters that were adjusted included: (1) The volume of the offered load, which was varied by manipulating the number of flows in the network. (2) The average node degree $d$, which represented the density of the network was set to 5, 9, 13 and 17 where 5 generated sparse networks and 17 generated dense networks. Table 2 summarizes the parameters that were used in simulations.

**Table 2. Some of simulation parameters.**

| Parameter | Value | Parameter | Value |
|---|---|---|---|
| Transmission power | 140 $\mu w$ | Transmission range ($\rho$) | 200 $m$ |
| Receive power | 90 $\mu w$ | Packet size | 500 $B$ |
| Idle listening power | 55 $\mu w$ | Number of nodes | 200 |
| Learning rate ($\beta$) | 0.5 | Traffic type | UDP |
| Discount factor ($\gamma$) | 0.9 | MAC protocol | 802.11 |
| Maximum feasible delay ($\theta_{max}$) | 8 | $c$ | 15 |

### 6.2. Simulation Results

Here, we evaluate the performance by simulating the proposed method in a multi-hop wireless network by concentrating on some of the more important plots (display in Figs 5, 6, 7, 8 and 9).

### 6.2.1. Convergence behavior

Our Q-learning algorithm, as any other learning scheme, needs a learning phase to learn the optimal decision policies. In [48] it was shown Q-learning convergence time was a polynomial function of the state and action size. To examine the convergence curve, the average obtained reward as a function of time is displayed in Fig. 5. The average convergence time as can be seen is around 4000 learning iterations. In this scenario, the parameters were set to: d=5, as average node degree and offered network load= 900kbps.

### 6.2.2. Average Coding Gain

We defined the coding gain as the ratio of the number of transmissions required by the non-coding approach to the minimum number of transmissions used by COPE or the proposed scheme to deliver the same set of packets to the next-hops successfully [52]. Fig. 6 compares the proposed coding gain against that of COPE under various conditions. When the number of flows increases in the network, the coding opportunities increases too. As can be seen in Fig 6, at lower load in the network, the coding gain behavior in both algorithms is similar, but when the offered load is increased, though the gain increases in both cases, the COPE model outperforms (by about 2.5%) which is due to some probable missed beneficial packets that node could not overhear because of sleep decisions in proposed scheme. In this setup, average node degree was set to 5.

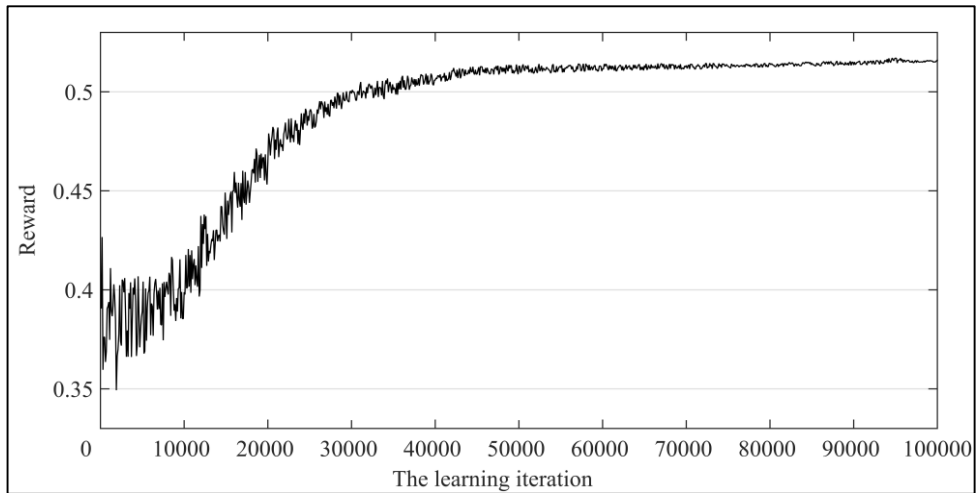
**Fig. 5. Convergence curve.**

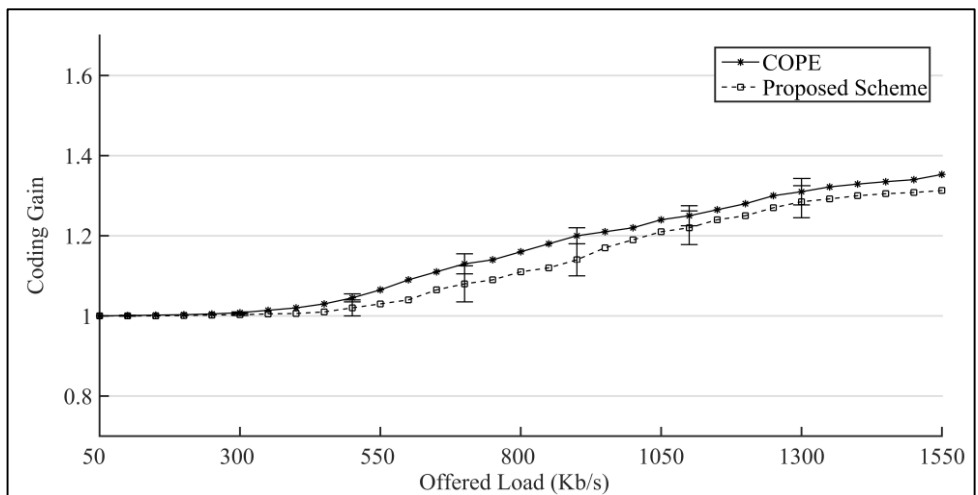
**Fig. 6. Coding gain vs. offered load**

### 6.2.3. Energy Consumption

We computed the average energy consumption per bit by dividing the total energy expenditure (over all the nodes in the network) by the total number of bits that delivered by final destinations. By increasing the offered load, the average energy consumed per bit increased smoothly. As shown in Fig. 7, in the higher traffic load environment, the average energy consumed per bit increased with higher rate due to congestion and collisions. What is important to note is that in the proposed scheme nodes sleep over through some of idle periods, hence they save energy rather than overhear irrelevant packets. In this scenario, $d$ was set to 5, as average node degree. In comparison, the consumption per bit in the COPE scheme is higher due to non-stop overhearing.

Fig. 8 illustrates the consumed energy per bit by the proposed solution versus that of the COPE scheme for changes in network densities, i.e. variation in the average node degree *d*. Network density does not noticeably affect consumption in the case of COPE, whereas it makes the proposed approach undergo significant reduction. After a detailed evaluation of the simulation results it can be concluded that the proposed algorithm offers a more efficient use of energy in more dense network environments because of more reduction in transmissions. In this scenario, offered load was set to 900kbps.

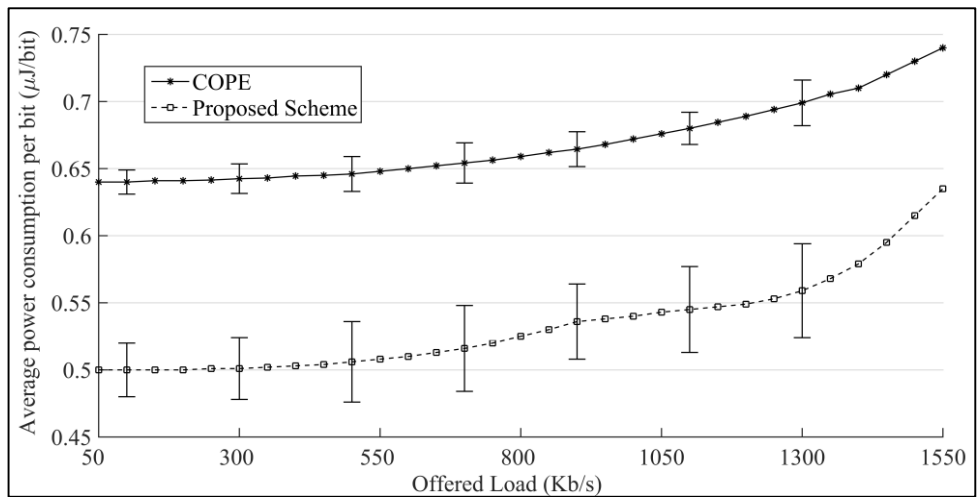

**Fig. 7. Energy consumption vs. offered load**

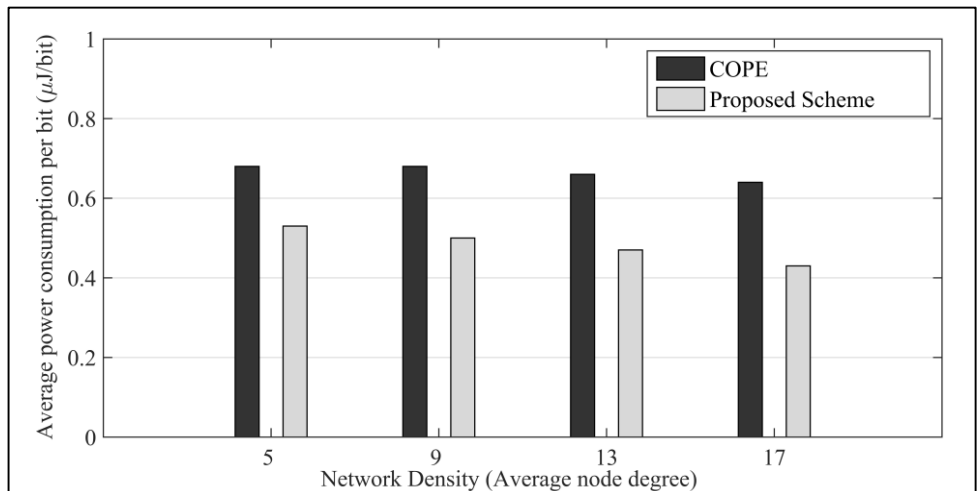

**Fig. 8. Energy consumption vs. network density**

**6.2.4. End-to-End Average Delay**

End-to-end delay includes all possible delays caused by buffering, queuing at the interface queue, scheduling at the MAC layer, propagation and transfer delays. It is the average time between the first transmission of a packet and the reception and successful decoding at the destination node. When the number of flows is small, both coding schemes perform similarly. But, as the number of flows increases the network congestion and wireless collisions increase with it and as a result end-to-end average delay rises dramatically. Fig. 9 shows delays in the proposed scheme is little longer than that in COPE. In particular, COPE operates better, which is due to better coding gain. In this scenario, the average coding degree was set to 5.

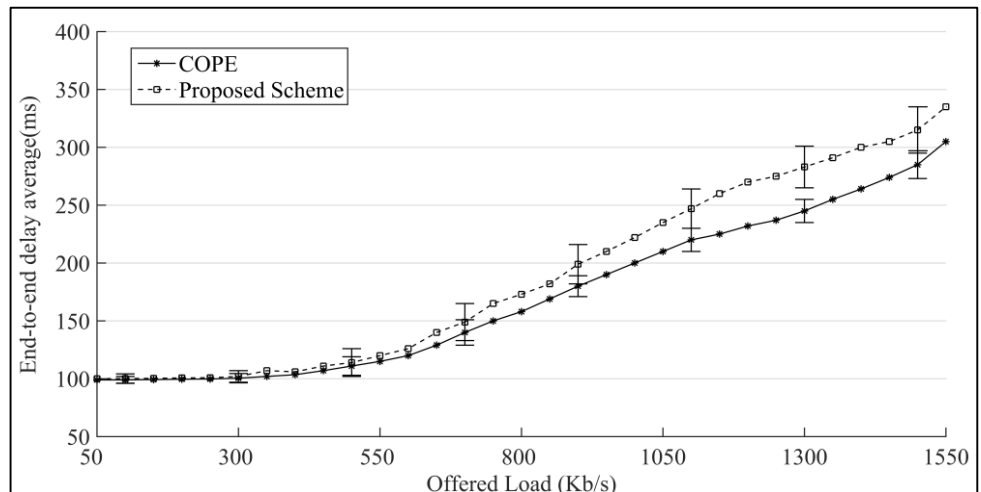

**Fig. 9. End-to-end delay vs. offered load**

## 7. Conclusion

In this paper, we considered adaptive sleep/wakeup scheduling for network coded multi-hop wireless networks. We know that there is a trade-off between energy consumption and network coding gain because some of overheard packets are useless for coding operation. And so, to determine our strategy for sleep/wakeup, we formulated SMDP model to decide whether to overhear or sleep. Thus, the node may turn off its transceiver during inactive periods in order to reduce energy consumption, or keep its transceiver on and remain awake in order to overhear more packets and consequently increase coding gain.

We proposed an on-line continuous-time learning method based on Q-learning. This approach enables the nodes to adapt their policies for the dynamic environment based on their local information and limited feedback from their neighboring nodes. In this proposal, a decision maker has to try to solve the problem resulting from delayed rewards. Simulation results show that our scheme improves energy consumption in multi-hop wireless networks significantly with negligible overhead in term of delay.